\documentstyle[11pt]{article}
\title{A POLYNOMIAL TIME ALGORITHM FOR LOCAL TESTABILITY AND ITS LEVEL}

\author{A.N. Trahtman}

\newtheorem{defn}{D\'efinition}[section]

\newtheorem{thm}[defn]{Theorem}

\begin{document}
\maketitle

\def\toeq{{@>\sim>>}} \def\into{{\hookrightarrow}}

\def\emptyset{{\varnothing}}

\def\alp{{\alpha}}  \def\bet{{\beta}} \def\gam{{\gamma}}
 \def\del{{\delta}}
\def\eps{{\varepsilon}}
\def\kap{{\kappa}}                   \def\Chi{{X}}
\def\lam{{\lambda}}
 \def\sig{{\sigma}}  \def\vphi{{\varphi}} \def\om{{\omega}}
\def\Gam{{\Gamma}}  \def\Del{{\Delta}}  \def\Sig{{\Sigma}}
\def\ups{{\upsilon}}

\def\Q{{\bf{Q}}}
\def\Z{{\bf{Z}}}
\def\R{{\bf R}}

\def\Gm{{\bf G}_{\rm{m}}}
\def\Gmk{{\bf G}_{{{\rm{m}},k}}}
\def\GmL{{\bf G}_{{{\rm{m}},L}}}
\def\GmK{{\bf G}_{{{\rm{m}},K}}}

\def\GmO{{\bf G}_{{{\rm{m}},{\cal O}}}}

\def\bW{{\bf W}}
\def\bG{{\bf G}}
\def\Ga{{\bf G}_{\rm{a}}}

\def\Fb{{\overline F}}
\def\Hb{{\overline H}}
\def\Yb{{\overline Y}}
\def\Xb{{\overline X}}
\def\TT{{\overline{T}}}
\def\Bb{{\overline B}}
\def\Gb{{\overline G}}
\def\kb{{\overline k}}
\def\TT0{{\overline T}^{\circ }}

\def\Th{{\hat T}}
\def\Bh{{\hat B}}

\def\cO{{\cal O}}
\def\cT{{\cal T}}
\def\cG{{\cal G}}
\def\cE{{\cal E}}
\def\cA{{\cal A}}
\def\cS{{\cal S}}
\def\T0{{\cT}^{\circ }}
\def\cR{{\cal R}}

\def\Xt{{\tilde X}}
\def\Gt{{\tilde G}}


\def\gg{{\Pi}}

\def\min{^{-1}}

\def\Cl{\rm{Cl}}            \def\Stab{{\rm{Stab}}}
\def\Gal{{\rm{Gal}}}          \def\Aut{\rm{Aut\,}}
\def\Lie{\rm{Lie\,}}
\def\GL{{\rm{GL}}}          \def\SL{{\rm{SL}}}
\def\loc{\rm{loc}}
\def\coker{\rm{coker\,}}    \def\Hom{{\rm{Hom}}}
\def\im{\rm{im\,}}           \def\int{\rm{int}}
\def\inv{\rm{inv}}           \def\can{\rm{can}}
\def\Spec{{\rm{Spec\,}}}

\def\min{{\rm{min}}}
\def\det{{\rm{det}}}

\def\tors{_{\rm{tors}}}      \def\tor{^{\rm{tor}}}
\def\red{^{\rm{red}}}         \def\nt{^{\rm{ssu}}}
\def\sssc{^{\rm{sssc}}}
\def\sss{^{\rm{ss}}}          \def\uu{^{\rm{u}}}
\def\ad{^{\rm{ad}}}           \def\mm{^{\rm{m}}}
\def\tm{^\times}                  \def\mult{^{\rm{mult}}}
\def\uss{^{\rm{ssu}}}         \def\ssu{^{\rm{ssu}}}
\def\cf{^{\rm{cf}}}
\def\ab{_{\rm{ab}}}

\def\til{\;\widetilde{}\;}

\def\emptyset{{\varnothing}}

\font\cyr=wncyr10 
\def\Bcyr{{\cyr B}}
\def\Sh{{\cyr Sh}}
\def\Ch{{\cyr Ch}}

\def\lpsi{{{}_\psi}}
\def\bks{{\backslash}}

\def\Pic{{\rm Pic\,\;}}
\def\Br{{\rm Br\,}}
\def\Brn{{\rm Br}_1\,}
\def\Brz{{\rm Br}_0\,}
\def\Bra{{\rm Br}_{{a}} \,}
\def\Brnr{{\rm Br}_{{nr}}}

\def\xb{{\overline x}}
\def\lc{{{}_c}}

\def\X{{\rm{X}}}



\centerline{International Journal of Algebra and Computation vol.9 no. 1(1998), 31-39}
\medskip

 Bar-Ilan University, Department of Mathematics and Computer Science,

 52900 Ramat Gan, ISRAEL

\medskip
{\it e-mail}: trakht@macs.biu.ac.il

\medskip

\maketitle

\begin{abstract} A locally testable semigroup  $S$  is a semigroup with
        the property that for some nonnegative integer  $k$, called the order
        or level of local testability, two words $u$ and $v$ in some
	set of generators for $S$
	are equal in the semigroup if  (1) the prefix and
        suffix of the words of length $k$ coincide, and (2) the set of
        intermediate
        substrings of length $k$ of the words coincide.
          The local testability problem for semigroups is, given a finite
        semigroup, to decide, if the semigroup is locally testable or not.

          Recently, we introduced a polynomial time algorithm for the local
        testability problem and to find the level of local testability
        for semigroups based on our previous description of identities
	of $k$-testable semigroups
        and the structure of locally testable semigroups.

          The first part of the algorithm we introduce solves the local
        testability problem.

          The second part of the algorithm finds the order of
        local testability of a semigroup. The algorithm is of order
        $n^2$, where $n$ is the order of the semigroup.
\end{abstract}


 \section{INTRODUCTION}.

   The concept of local testability was first introduced by McNaughton
 and Papert [5] and since then has been extensively investigated from
 different points of view (see [1]-[6], [8]). This concept is
 connected with languages, finite automata and semigroups. The purely
 algebraic approach proved to be fruitful (see [6], [8]) and in this paper
 we use this technique.

  The algorithms for the problem of local testability can be found in [1],
[5]. They are polynomial in terms of the size of the semigroup.
  In [3] a polynomial time algorithm for the local testability
 problem for a given deterministic finite automaton was given.
 The order of this algorithm is
 $sn^2$, where $n$ is the number of states of the automaton, and $s$ is the
 size of the alphabet.

 We introduce in this paper a new polynomial
 time algorithm for finding the level of local testability
 for a given semigroup.
 Necessary and sufficient conditions for a semigroup to be locally testable
 from [6] are used here.

 The solution
 is connected with the problem from [2] (see [1], [3] too):
 ``Is there a practical algorithm which, given a locally testable deterministic
 automaton, finds $k$ such that the automaton is properly $k$-testable (i.e.
 $k$-testable but not ($k$-1)-testable)?'' We give the answer to this question
 in the case that the automaton is the right regular representation of a
 semigroup $S$. In the general case computing the order of local
 testability of a locally testable automaton is NP-hard [4].

 The order of our
 algorithm is $n^2$, where $n$ is the order of the semigroup.
 The first part of the algorithm solves the local testability problem.
The second part finds the order of local testability.

 There are two different definitions of $k$-testability (see [2]). Our
 algorithm gives an answer in the sense of the definition used in [1] and [2].

 \section{NOTATION AND DEFINITIONS}

Let $\Sigma$ be an alphabet and let $\Sigma^{+}$ denote the free semigroup
on $\Sigma$. If $w \in \Sigma^{+}$, let $|w|$ denote the length of $w$. Let
$k$ be a positive integer. Let $i_{k}(w) (t_{k}(w))$ denote the prefix
(suffix) of $w$ of length $k$ or $w$ if $|w| < k$. Let $F_{k}(w)$ denote the
set of factors of $w$ of length $k$. That is,
$F_{k}(w) = \{x \in \Sigma^{+} | |x| = k$ and $ w = uxv$ for some $u,v \in
\Sigma^{+}\}$.

A semigroup $S$ is called {\em k-testable} if there is an alphabet $\Sigma$
and a surjective morphism $\phi:\Sigma^{+} \rightarrow S$ such that for all
$u,v \in \Sigma^{+}$, if $ i_{k}(u) = i_{k}(v), t_{k}(u) = t_{k}(v)$ and
$F_{k}(u) = F_{k}(v)$, then $u\phi\ = v\phi$.
This definition follows [4], [5], [6], [8], but [2] and [1].
In [1] the definition differs by considering prefixes and suffixes of
length $k$-1.

A semigroup $S$ is {\it locally testable} if it is $k$-testable for some
$k$.
For local testability the two definitions mentioned above are equivalent [2].

It is known that the set of $k$-testable semigroups forms a variety of
semigroups [8].
Let $T_k$ denote the variety of $k$-testable semigroups.

We need the following notation and definitions.

$|S|$ is the number of elements of the set $S$.

$S^m$ denotes the ideal of the semigroup $S$ containing products of elements
 of $S$ of length $m$ and greater.

[$u=v$] denotes the variety of semigroups defined by identity $u=v$.
By definition, $S \in [u=v$] if and only if the identity $u=v$  holds in $S$.

A semigroup $S \in [xy=x]$ [$S \in [xy=y]$] is called {\em semigroup of left
[right] zeroes}.

We will need to consider the following semigroup
$A_2=<a,b| aba=a, bab=b, aa=a, bb=0>$.
S is a 5-element 0-simple semigroup, $A_2=\{a, b, ab, ba, 0 \}$, in which
only $b$ is not an idempotent. The basis of identities $A_2$ is the
following [7].

\begin{equation}
x^2 = x^3, xyx=xyxyx, xyxzx=xzxyx.          \label{(1)}
\end{equation}

According to [6] $T_n$ has the following basis of identities:

\begin{equation}
\alpha_r: (x_1 ... x_r)^{m+1}x_1...x_p=(x_1...x_r)^{m+2}x_1...x_p 
\label{(2)}, \end{equation}

where $r \in \{1,...n\}$, $p=n-1(mod$ $r), m=(n-p-1)/r$,

\begin{equation}
\beta: t_1x_1...x_{n-1}yx_1...x_{n-1}zx_1...x_{n-1}t_2=%
t_1x_1...x_{n-1}zx_1...x_{n-1}yx_1...x_{n-1}t_2.   \label{(3)}
\end{equation}
For instance, $\alpha_1$ is the identity: $x^n=x^{n+1}$. A locally testable
 semigroup $S$ has only  trivial subgroups [5] and so a locally testable
 semigroup $S$ with $n$
elements  satisfies  identity $\alpha_1$.
\medskip
\section{SOME AUXILIARY RESULTS}

  \proclaim Lemma 1. Let a semigroup $S$ be locally testable and assume that
$S^k=S^{k+1}$.
Then the ideal $S^k$ is 2-testable and belongs to $var A_2$.

    Proof. A locally testable semigroup $S$ satisfies identities (2), (3) for
  some $n$ and for all numbers greater than $n$. So we may  suppose that
  $n \geq k$. All words from $S^k$ may be presented as words of length $k$
  and greater and so $S^n=S^k$. The identity $\alpha_{k-1}$:
($x_1...x_k)^2=(x_1...x_k)^3$ of
  $S$ implies the identity $x^2=x^3$ in $S^k$. The identity $\alpha_k$ of
$S$ implies the identity  $xyx=xyxyx$ in $S^k$. Now consider a word $a$
from $S^k$. The word $a$ may be presented in the form $a=t_1bt_2$,
where $b$ is of length $k$. Then
             $$ayaza=t_1bt_2yt_1bt_2zt_1bt_2$$.
  Using identity (3) for the words $b$, $t_2yt_1$, $t_2zt_1$ we see that

      $ayaza=t_1bt_2yt_1bt_2zt_1bt_2=t_1bt_2zt_1bt_2yt_1bt_2=azaya$.

 So all the identities  (1) are true in $S^k$. Thus $S^k$ belongs to $var
  A_2$ and is
  2-testable because $A_2$ is 2-testable.

 The lemma is proved.\medskip

    Now from the necessary and sufficient conditions of local testability [6]      we have the following.

 \proclaim Corollary.
 Let $S^n=S^{n+1}$ for some $n$ in some semigroup $S$.
 Then $S$ is locally testable iff $S^n$ belongs to $var A_2$.
  \medskip

   The following statement is well known.
 \proclaim Lemma 2.
 Let $S$ be a finite semigroup, $S^m=S^{m+1}$ for some $m$.
 Let $E$ be the set of idempotents of $S$.
 Then any element of $S^m$ is divided by an idempotent, that is  $S^m = SES$.
\medskip

    Proof. Let $a$ belong to $S^m$. Then $a$ may be represented by a word of
 length greater then $| S |$. So we can construct a chain of left subwords
 of $a$ such that each element of the chain is a left divisor of the following
 element and the number of elements is greater then $| S |$. This implies that
 there are two different left subwords $b$ and $bc$ such that $b=bc$. Then
 $b=bc=bcc$ and $b=bc^n$. For some  $n$, $c^n$ is an idempotent and a
 right unit of $b$. The element $b$ divides $a$ and the right unit of $b$
 divides $a$ too. Thus $S^m$ is contained in $SES$. The opposite inclusion
 is obvious. So $S^m=SES$.
  \medskip

 {\bf Lemma 3.}
Let $S$ be a finite semigroup such that
$S^2=S$ and $S \in [x^2=x^3, xyx=xyxyx$].

The following two conditions are equivalent in $S$:

{\bf a)} $S$ satisfies the identity  $xyxzx=xzxyx$.

{\bf b)} No two distinct idempotents $e$, $i$ from $S$ such
that  $eie=e, iei=i$ have a common unit in $S$. That is, there is no
idempotent $f \in S$ such that $ef = e = fe$ and $if = i = fi$.

    Proof Suppose $S$  belongs to [$xzxyx=xyxzx$] and for some idempotents
 $e$, $i$ in $S$ $eie=e$, $iei=i$. Suppose $f$ is a common unit of $e$, $i$.
 The identity $xyxzx=xzxyx$ implies that $ei=fefif=fifef=ie$.
  Now $e=eie=eei=ei=eii=iei=i$. Thus the idempotents $e$, $i$ are not distinct.

    Suppose now that $S$ does not belong to [$xyxzx=xzxyx$]. So for some
 $a$, $b$, $c$ of $S$  $abaca \neq acaba$. Since  $S^2=S$, Lemma 2
 implies that $a$ is divided by some idempotent $e$, $a=peq$. Then
 $peqbpeqcpeq \neq peqcpeqbpeq$. This implies that $eqbpeqcpe$
 and $eqcpeqbpe$ are distinct. From the identity $xyx=xyxyx$
 it follows that $i=eqbpe$  and $j=eqcpe$  are idempotents. They have the
common unit $e$ and are distinct, because $ij$ and $ji$ are distinct.

   Consider the elements $iji$ and $jij$. In view of the identity $xyx=xyxyx$
 they are idempotents too. It is routine to prove that they belong to
an idempotent subsemigroup of $S$.
The element $e$ is then a common unit for $iji$ and $jij$.

 Now suppose that $iji=jij$.
 We have  $ij=eije=eijeije=eijije=ejijje=jij$. Analogously
 $ji=jij$. So $ij=ji$ in contradiction to the above result. We conclude
 that $iji \neq jij$. So the distinct idempotents $iji$, $jij$ belong to
 a band and have common unit $e$.

The lemma is proved.

  From the Corollary to lemma 1 and lemmas 2 and 3 we
  obtain the following result.
\begin{thm}  Let  $S$ be a finite semigroup, and let $E$ be the subset of
 idempotents of $S$.
    Suppose that $SES$ satisfies the
    identities $x^2=x^3$, $xyx=xyxyx$ and every two idempotents $i$, $j$
in $S$ having a common unit and such that $ij=i$, $ji=j$ or $ij=j$, $ji=i$
 coincide. Then $S$ is locally testable.
 \end{thm}  \medskip
    This theorem will be the basis for the first part of the
  algorithm to
    verify the
 local testability of a finite semigroup.

    Recall that a semigroup $S$ is called {\em locally idempotent} iff
$eSe$ is an idempotent subsemigroup for any idempotent $e \in S$.
Obviously, the set $SES$ for a locally idempotent semigroup $S$ with
set of idempotents $E$
 satisfies identities $x^2=x^3$, $xyx=xyxyx$. A locally testable semigroup is
 locally idempotent [1], [5].
 Then from the
result of the last theorem follows.

 \begin{thm} [1] A finite semigroup $S$ is locally testable iff it is locally
idempotent and $S$ does not contain the three-element monoid with two left
[right] zeroes. That is, $S$ is locally testable iff $eSe$ is a semilattice
for all $e=e^2 \in S$.
\end {thm}  \medskip

    Now we consider the definition of $n$-testability from [1] and [2]. The
 results of [8] and [6] may be repeated in this case too. We present this
 fact without proof. Consider the following identity.

\begin{equation}
x_1...x_{n-1}yx_1...x_{n-1}zx_1...x_{n-1}=x_1...x_{n-1}zx_1...x_{n-1}yx_1...x_{n-1}.
 \label{(4)}
\end{equation}

Let $B_n$ be the set of $n$-testable semigroups in the sense of [1].

 \begin{thm}

   a) $B_n$ is a variety,

   b) A basis of identities for $B_n$ for $n \geq 2$ consists of
 identities (2) and the identity (4),

   c) $B_1$=[$x^2=x, xy=yx$].

 \end{thm}  \medskip

   The only difference between the identities (3) and (4) is the omission in (4)
of the first and last letters $t$ from (3). Therefore, we have the following.

 \proclaim Corollary. $T_n$ contains $B_n$. $B_n$ contains $T_{n-1}$.
 $B_2=varA_2$.

 The following lemma enables us to find the level of local
 testability of a semigroup $S$.

 {\bf Lemma 4.}
Let $S$ be a finite semigroup satisfying the identities (2)
 for some $n$. Then the following two conditions are equivalent in $S$:

{\bf a)} The semigroup $S$ is $n$-testable.

{\bf b)} Every two distinct idempotents $e$, $i$ in $S$ such that
 $ei=e, ie=i$  [$ei=i,ie=e$] have no common left [right] divisor in
 $S^{n-1}$.

    Proof: Let us denote $X=x_1...x_{n-1}$.

    Suppose first that $S$ is $n$-testable. Then
 according to theorem 3.3, $S$ satisfies the identity

      \begin{equation} XyXzX=XzXyX.               \label{(5)}
      \end{equation}

 Consider idempotents $e$, $i$ such that $ie=i,ei=e$ having common
left divisor  $a$ in $S^{n-1}$. We will prove that $e=i$. Let $i=ab$,
$e=ac$. We have  $i=ie=iei=abacab$. The identity (5) implies that
$abacab=acabab=eii=ei=e$. So  $e=i$, and a) implies b).

 Suppose now that $S$ is not $n$-testable. Then $S$ does not satisfy identity
 (5). So for some elements $a$, $b$, $c$, where $a \in S^{n-1}$, we have
 $abaca \neq acaba$. One of the equalities $abaca=abacaba$,
 $acaba=abacaba$  does not hold in $S$. Without loss of generality suppose
 that $acaba$ and $abacaba$ are distinct. In view of the identity

   \begin{equation}              XyX=XyXyX, \label{(6)}
   \end{equation}

 we have
 $abacaba \neq acabacaba$. Then $abacab$ and $acabacab$ are distinct.
 Let as denote $e$=$abacab$, $i$=$acabacab$. $e$, $i$ are distinct. Using (6)
 we have

 $e=abacab=abacabacab=abacababacab$=$e^2$. Analogously $i$ is
 idempotent too. Now $ei$=$abacabacabacab$=$abacab$=$e$. Analogously
 $ie$=$i$. So we find two distinct left zeroes having a common left divisor
 in $S^{n-1}$.

 \medskip

Let $S$ be a finite locally testable semigroup and let
$\phi:\Sigma^{+} \rightarrow S$ be a surjective morphism for some alphabet
$\Sigma$. Let $a$ be an element from $S \setminus SES$.
Let $m$ be  the maximal number such that $a^{m+1} \neq a^{m+2}$.
 Suppose $a = bc$.
Since $a$ belongs to $S \setminus SES$, it follows that  $a, b, c$
have only a finite number of preimages in $\Sigma^{+}$.
Denote $|a|$, $|b|$, $|c|$ the maximal length of the preimages
of the elements $a$, $b$, $c$ in the alphabet $\Sigma$, correspondingly.
Suppose $n=max((|b|+|c|)m+|b|+1$ if $a^{m+1}b \neq a^{m+2}b, |a|m+1$ otherwise)
 for all $a \in S \setminus SES$ and for all $b,c$ such that $bc=a$.

\proclaim Lemma 5.
The minimal number for which $S$ satisfies identities (2) is equal to
$n$+1.
\medskip

Proof: Consider some identity $\alpha_r$ from (2) and the
corresponding words $(a_1...a_r)^{m+1}a_1...a_p$,
$(a_1...a_r)^{m+2}a_1...a_p$.Denote $a=a_1...a_r$, $b=a_1...a_p$,
$c=a_{p+1}...a_r$. Because $a^{m+1} \neq a^{m+2}$ the semigroup $S$
is not $n$-testable for $n=m|a|$+1.
If the words $a^{m+1}b$ and $a^{m+2}b$ are not equal then the semigroup
$S$ is not $n$-testable for
$n=m$ ($|b| + |c|$)+$|b|$+1. So the maximum of all such
numbers $n$ gives us the precise bound for which the
identities (2) are not valid.

From Lemmas 4 and 5 we have the following theorem.

\begin{thm} \label{3}. Let $S$ be a finite locally testable semigroup. Let
 $a$ be an element of $S \setminus SES$. Let $b$ be a proper left divisor of
 $a$ and $a=bc$.
 Let $|a|$, $|b|$, $|c|$ be the maximal length of the words $a,b,c$ in some
 alphabet $\Sigma$.
 Let $m$ be the maximal number such that
 $a^{m+1} \neq a^{m+2}$.
 Let $n$=max(($|b|+|c|)m+|b|+1$ if $a^{m+1}b \neq a^{m+2}b, |a|m$+1 otherwise)
 for all $a \in S \setminus SES$  and for all  $b,c$ such that $bc=a$.

   Let $e$, $i$ be idempotents of a left [right] zero subsemigroup and let
 $a$ be a left [right] common divisor of maximal length. Let
 $l(e,i)=| a | +1$ [$r(e,i)=| a | +1$] and let $l$=max($l(e,i)$)
 [$r$=max($r(e,i)$)] for all pairs of left [right] zeroes.

   Then max($n,r,l$)+1 is equal to the exact level of local testability of the
 semigroup $S$.
 \end{thm} \medskip

Recall that a semigroup $S$ is a left (right) zero semigroup if $S$
satisfies the identity $xy = x (xy = y)$.
The following proposition is useful for the next algorithm.

{\bf Proposition}
Let $E$ be the set of idempotents of a semigroup and let
$|E|=n$. We represent $E$ as an ordered list $[e_{1}, \ldots , e_{n}$].
Then there exists an algorithm of order $n^2$ that
reorders the list so that the maximal left (right) zero subsemigroups of $S$
appear consecutively in the list.

 \medskip

Finding the maximal semigroup of left zeroes containing a given
idempotent needs $n$ steps. So to reorder $E$ we need at most $n^2$ steps.

 \section{ALGORITHMS}

   1.{\it Testing whether a finite semigroup $S$ is locally testable.}
 \medskip

 Suppose $| S | = k$.
 We begin by finding the set of idempotents $E$.
 This is a linear time algorithm.
 After this we find $SE$ and then $SES$ using
 two times $O(k^2)$ steps.

   In view of Theorem 3.1 we begin by verifying the first two identities from (1)
 in $SES$. Verifying the identity $x^2=x^3$ needs $O(k)$ steps, verifying the
 identity $xyx=xyxyx$ needs $O(k^2)$ steps.

   Now consider the last identity from (1). In view of Lemma 3 consider
 the set $E$. We can reorder $E$ according to the proposition above in a
 chain such that the subsemigroups of left zeroes form intervals in this
 chain. We note the bounds of these
 intervals. We find for each element $e$ of $E$ the first element $i$ in
 the chain such that $e$ is a unit for $i$. Then we find in the chain
 the next element $j$ with the same unit $e$. If $i$ and $j$ belong
 to the same subsemigroup of left zeroes we conclude that $S$ is not testable
 (Lemma 3) and end the process. If they are in different left zero semigroups,
 we replace $i$ by $j$ and continue the process of finding a new $j$. This
 takes  $O(k^2)$  steps.

   Then we repeat the same process for right zeroes. According to Theorem 3.1 we can
 give a positive answer to the question in the case we do not find two
 different left [right] idempotents with the same unit.

 \medskip

  2.{\it Finding the level of local testability.}
\medskip

  The idea is based on theorem 3.4.

  Suppose the semigroup $S$ is locally testable, $| S| =k$, $E$ is the
 set of idempotents of $S$, as above.

 We use the sets $E$, $SE$ and $SES$ found above. According to lemma 1,
 $SES=S^l$, where $S^l=S^{l+1}$. We find $G=S \setminus SES$ as well.

   In the case that $G$ is empty the semigroup $S$ belongs to $varA_2$ and is
 2-testable in both senses. Verifying of 1-testability reduces to testing
 the identities $x=x^2$, $xy=yx$. So we are done in this case.
  Now suppose that $G$ is not empty.

   We start with the assignment $n:=2$ to a variable $n$ that will change
   during the algorithm below.

   For each element $a$ we find a maximal number $m$ that
 $a^{m+1} \neq a^{m+2}$.
 Denote this number by $m(a)$.
 This is an algorithm of order at most $k$.
 The maximum of all such $m(a)$ gives us a first lower bound of the order of
 local testability.

   Then we find generators for $G$. It is easy to see that the unique
 minimal generating set for $G$ is $G \setminus G^2$. This takes
 $k^2$ steps.
 We denote the set of these elements by $G_1$.
The  maximal length of elements from $G_1$ is
 1. We now define a sequence of sets $G_{i}, i \geq 1$. We want $G_{i}$ to be
 equal to the set of elements in $G$ that can be written as a product of $i$
 elements from $G_{1}$, but that cannot be written as a product of
 more than $i$
 elements of $G_{1}$. Assume that we have correctly defined $G_{i}$ for some
 $i \geq 1$. Then we let
 $G_{i+1} = (G_{1}G_{i})\setminus ((G\setminus G_{1})G_{i})$. It is
 easy to see by induction that this correctly defines $G_{i}$ for $1 \leq i
 \leq l-1$. Elements of $G_{i}$ are said to have level $i$.
 Each element of $G$ has a well defined level. This process need $k^2$ steps,
 because  $|G| =|G_1|+...|G_{l-1}|$.
  We know that the level of an element $g \in G$ is equal to its
 maximal length in any set of generators for $G$ and will be
 denoted by $|g|$.

 Consider now all possible products $bc$ for $b$, $c$ from $G$.
 This takes $O(k^2)$ and on each step for $a \in G$ we do the following:

   Suppose that $a=bc$. Let $n = max(n, m(a)|a| + 1)$.
 Consider the element
 $a^{m+2}b$. If this is not equal to the element $a^{m+1}b$, we make the
 following assignment $n:=max(n, (|b|+|c|)m+|b|+1)$.

   After considering all pairs we get a value for $n$.
 In view of Lemma 5 the semigroup
 $S$ does not satisfy identities (2) for $n$ and satisfies (2) for $n$+1.

   Now consider the identity (4). What follows will be based on
 Lemma 4.
  We first reorder the set of idempotents $E$ as in the proposition for
 left zero subsemigroups. We note the bounds between the subsemigroups.
 In view of the Proposition this takes at most $O(k^2)$ steps.

 Let us assign $L := n$.

  For each $g$ of $G$ we form the intersection $gSES \cap E$. Then we
 verify: are all elements of the intersection within the bounds or not. If
 there are two idempotents not within the bounds the
 semigroup $S$ is not ($|g|+1$)-testable and may be only
  $| g |+2$-testable. Now assign $L:=max(L, |g|+1)$.
 This needs at most 2$k$ steps. Repeating this
 process for all $g$ from $G$, we find the maximum $L$ for all such $g$, using
 at most $2k^2$ steps. The semigroup $S$ is not $L$-testable and may be
 only $L+1$-testable.

   Then we repeat the procedure for the right order of $E$ and right divisors
 of idempotents. As a result the upper bound $R$ may be obtained.

   Theorem 3.3 gives us the level of local testability as 1 + maximum of the
 three above-mentioned numbers $n$, $R$ and $L$.

  The algorithm is a polynomial time algorithm of order $O(k^2)$, where $k$ is
the order of the semigroup.

  Both parts of the algorithm give us a way to verify testability and to
 find its level.
\medskip

\section{ACKNOWLEDGMENT}

  The author thanks Stuart Margolis for helpful and fruitful discussion.

 \end{document}